\documentclass[twoside,twocolumn,english,prc,amsmath]{revtex4}
\usepackage[T1]{fontenc}
\usepackage{amsmath}
\usepackage{graphicx}
\usepackage{amssymb}

\makeatletter
\def\@oddhead{\rightmark \hfill A Realistic Treatment of Geomagnetic Cherenkov Radiation  \hfill \thepage}
\def\@evenhead{\thepage \hfill K. Werner, K.D. de Vries, O. Scholten \hfill}
\def\fnum@table{\tablename~{\bf\thetable}}
\def\fnum@figure{\figurename~{\bf\thefigure}}
\def\tablename{\footnotesize{\bf Table}}
\def\figurename{\footnotesize{\bf Figure}}

\def\VYP#1#2#3{{\bf #1}, #3 (#2)}  
\newcommand{\etal}{\mbox{\textit et al.}}                       %
\usepackage{dcolumn}

\def\citet{\cite}

\makeatother

\usepackage{babel}

\begin{document}

\title{A Realistic Treatment of Geomagnetic Cherenkov Radiation \\
from Cosmic Ray Air Showers }

\author{\textbf{Klaus WERNER }}

\affiliation{SUBATECH, Université de Nantes -- IN2P3/CNRS -- EMN, Nantes, France}

\author{\textbf{Krijn D. DE VRIES, Olaf SCHOLTEN }}

\affiliation{Kernfysisch Versneller Instituut, University of Groningen,9747 AA,
Groningen, The Netherlands}

\begin{abstract}
We present a macroscopic calculation of coherent electro-magnetic
radiation from air showers initiated by ultra-high energy cosmic rays,
based on currents obtained from three-dimensional Monte Carlo simulations
of air showers in a realistic geo-magnetic field. We discuss the importance
of a correct treatment of the index of refraction in air, given by
the law of Gladstone and Dale, which affects the pulses enormously
for certain configurations, compared to a simplified treatment using
a constant index. We predict in particular a geomagnetic Cherenkov
radiation, which provides strong signals at high frequencies (GHz),
for certain geometries together with {}``normal radiation'' from
the shower maximum, leading to a double peak structure in the frequency
spectrum. We also provide some information about the numerical procedures
referred to as EVA 1.0.
\end{abstract}
\maketitle

\section{Introduction}

The aim of our work is to provide a realistic calculation of radio
emission from air showers, which might be used finally to analyze
and understand the results from radio detection experiments ({\small LOPES}~\citet{Fal05,Ape06},
{\small CODALEMA}~\citet{Ard06}, LOFAR~\citet{lofar2}), and the
new set-ups at the Pierre Auger Observatory (MAXIMA \citet{auger1},
AERA \citet{auger2}). 

There are two ways to compute the electric fields created by the moving
charges of air showers: the {}``macroscopic approach'' adds up the
elementary charges and currents to obtain a macroscopic description
of the total electric current in space and time, which is the source
of the electric field obtained from solving Mawell's equations. The
{}``microscopic approach'' computes the fields for each elementary
charge, and adds then all the fields (with a large amount of cancellations). 

Already in the earliest works of ~\citet{Jel65,Por65,Kah66,All71},
a macroscopic treatment of the radio emission was proposed, but at
the time the assumptions about the currents were rather crude. There
is recent progress concerning the macroscopic approach. In 2007, we
performed calculations allowing under simplifying conditions to obtain
a simple analytic expression for the pulse shape, showing a clear
relation between the pulse shape and the shower profile \citet{olaf}.
This allows, for example, to determine from the radio signal the chemical
composition \citet{krijn2} of the cosmic ray. The picture used was
very similar to the one in Ref.~\citet{Kah66}, which has been refined
by using a more realistic shower profile and where  we calculate the
time-dependence of the pulse. Recently it was confirmed that the pulse
predicted in the microscopic description \citet{huegear1,alvarez}
agrees with the predictions following from the macroscopic picture
as shown in \citet{huegear2}.

In Ref.~\citet{klaus1}, we advance further by computing first the
four-current from a realistic Monte Carlo simulation (in the presence
of a geo-magnetic field), and then solve the Maxwell equations to
obtain the electric field, while considering a realistic (variable)
index of refraction, given by the Gladstone-Dale law as\begin{equation}
n=n_{GD}=1+0.226\frac{\mathrm{cm^{3}}}{g}\rho(h),\end{equation}
with $\rho(h)$ being the density of air at an atmospheric height
$h$. Although this index varies only between 1 and 1.0003, this variation
has important consequences, as discussed in detail in Ref.~\citet{klaus1}.
For example, the retarded time $t^{*}$ (the time when the signal
was sent out) for a given observer position is a multivalued function
of the observer time $t$, which gives rise to {}``Cherenkov effect''
phenomena, where the signal may become very short and very strong.
The caveat in this treatment is the fact that we consider the currents
to be point-like, which is only a good approximation far from the
shower axis. The Cherenkov-like effects actually show up as singularities,
and we expect these singularities to disappear when we give up the
{}``point-like'' assumption. Nevertheless, although Ref.~\citet{klaus1}
does not provide a realistic picture for all observer distances, its
results are very important as the basis of the much more realistic
description employed in the present paper.

\begin{figure*}[tbh]
\includegraphics[scale=0.8]{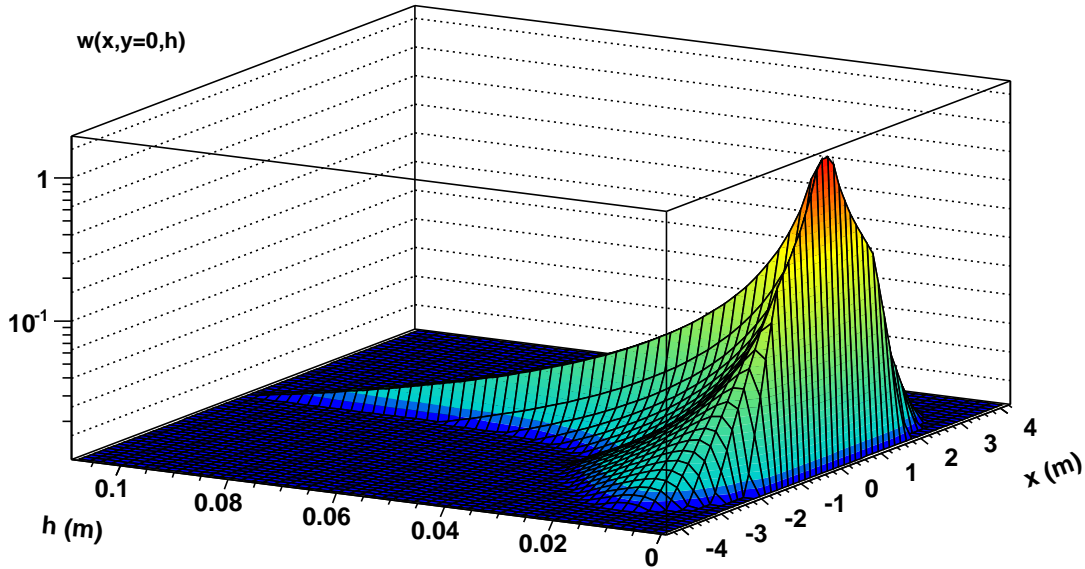}

\caption{The distribution of charged particles $w(x,y,h)$ at a given time,
as a function of the transverse coordinate $x$ and the longitudinal
coordinate $h$, for $y=0$.\label{fig:www}}

\end{figure*}
Anyhow, the notion {}``point-like'' has to be taken with care. In
the point-like picture described in Ref.~\citet{klaus1}, we do not
have a simple moving point-like charge, we rather have already transverse
currents, and also the longitudinal structure is nontrivial, just
all these currents are -- at a given time -- concentrated in a very
small volume. But there must be an internal structure, and therefore
it is natural as a next step to investigate the three-dimensional
structure of the shower at a given time. In order to do so, we consider
a {}``shower fixed'' coordinate system. The origin $O$ of this
system is the center of the shower front. We use the coordinates $x$
and $y$ to describe positions in the plane transverse to the shower
axis, and $h$ as the longitudinal distance behind the shower front.
The latter one is actually a hypothetical plane, which contains real
particles only around $x=y=0$, whereas for larger distances, the
fastest particles stay behind this plane. The situation as obtained
in a realistic Monte Carlo simulation (details to be discussed later)
is shown in fig. \ref{fig:www}. The distribution of charged particles
shows a very sharp maximum at the origin ($x=y=h=0)$, and falls steeply
in transverse and longitudinal direction. We will discuss the functional
form of this distribution later in detail, for the moment we only
want to illustrate the fact that the distribution obtained from simulations
shows nontrivial structures, concentrated in a small range in particular
concerning the $h$ variable. 

In the current paper, we want to take into account the realistic three-dimensional
form (at a given time) of the shower, as obtained from shower simulations,
still using a realistic index of refraction. The numerical procedures
of our approach, referred to as EVA$\,$1 (\underbar{E}lectric fields,
using a \underbar{V}ariable index of refraction in \underbar{A}ir
shower simulations), amount to air shower simulations, analysis tools
for extracting currents and shower shapes, and automatic fitting procedures
providing smooth functions for all relevant shower characteristics.
First results of our new approach have been published recently \citet{krijn}.
In the last part of the paper, we discuss important consequences of
our approach, referred to as {}``geomagnetic Cherenkov radiation'',
which provides strong emissions in the GHz frequency domain, alone
or as double peak structures in the frequency spectrum.

\section{Taming singularities}

We first repeat some elementary facts of the shower evolution, which
have been discussed in detail in Ref.~\citet{klaus1}. We consider
here showers due to a very energetic primary particle, with an energy
above $10^{14}\,\mathrm{eV}$. Such a shower moves with a velocity
$\beta c$, which is very close to the vacuum velocity of light $c$.
There is a constant creation of electrons and positrons at the shower
front, with somewhat more electrons than positrons (electron excess).
This is compensated by positive ions in the air, essentially at rest.
The electrons and positrons of the shower scatter and lose energy,
and therefore they move slower than the shower front, falling behind,
and finally drop out as {}``slow electrons / positrons''. Close
to the shower maximum, the charge excess of the {}``dropping out''
particles is compensated by the positive ions, since there is no current
before or behind the shower. Taking all together we have the situation
of a moving charge, moving with the vacuum velocity of light, even
though the electrons and positrons are moving slower, and they are
deviated (in opposite directions) in the Earth magnetic field.

\begin{figure*}[tbh]
\includegraphics[clip,angle=270,scale=0.4]{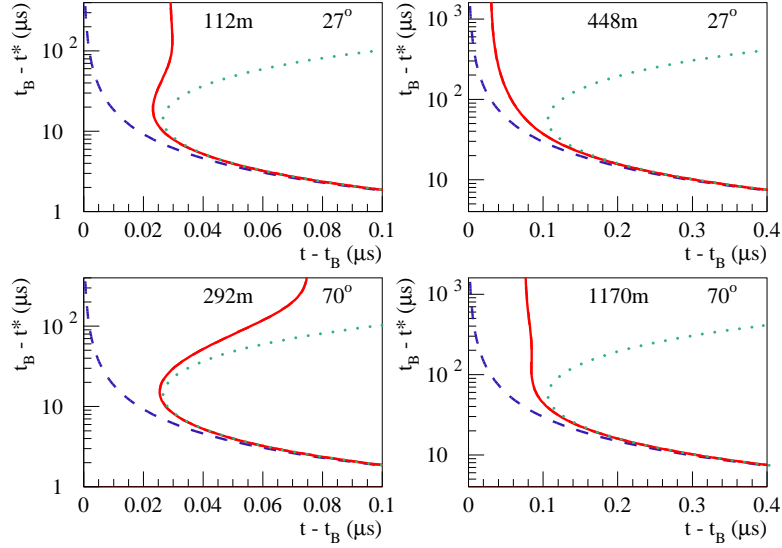}

\caption{The dependence of the retarded time $t^{*}$ on the observer time
$t$ for $n=1$ (dashed line), $n=n_{GD}$ (solid line), and $n=n_{\mathrm{ground}}\approx1.0003$
(dotted line) for inclined showers (27$^{o}$ and 70$^{o}$ and for
different distances in meters of the observer from the impact). \label{fig:trett1}
The reference time $t_{B}$ is the time of closest approach of the
shower with respect to the observer.}

\end{figure*}
Neglecting the finite dimension of the shower, referred to as {}``point-like''
(PL) approximation, one has a four-current \begin{equation}
j_{\mathrm{PL}}(t',\vec{x})=J(t')\,\delta^{3}(\vec{x}-\vec{\xi}(t')),\label{eq:current}\end{equation}
with a longitudinal component due to charge excess, and a transverse
component due to the geo-magnetic field. Solving Maxwell's equations,
we can express the potential in terms of the four-current $J$, evaluated
at the retarded time $t^{*}$, as \citet{klaus1} \begin{equation}
A_{\mathrm{PL}}^{\beta}(t,\vec{x})=\frac{\mu_{0}}{4\pi}\,\frac{J^{\beta}}{|\widetilde{R}V|}\,,\end{equation}
with $V=c^{-1}d\xi/dt'$, and with $\widetilde{R}$ being a four-vector
defined as $\widetilde{R}{}^{0}=c(t-t^{*})$ and $\widetilde{R}{}^{i}=-L\partial/\partial\xi^{i}L,$
where $L$ is the optical path length between the source $\vec{\xi}(t^{*})$
and the observer. This point-like approximation is certainly only
valid at large impact parameters ($>500\,\mathrm{m}$), but even more
importantly it will serve as a basis for more realistic calculations,
as discussed later. It should be noted that in case of $n>1$ and
even more for $n=n_{GD}$ the vector potential shows singularities,
which arise from $1/|\widetilde{R}V|\propto$ $dt^{*}/dt$ and the
fact that $t^{*}$ is a non-monotonic function of $t$, as shown in
fig. \ref{fig:trett1} and discussed in detail in \citet{klaus1}.
We show the realistic case $n=n_{GD}$ with the corresponding curve
situated between the two limiting cases $n=1$ and $n=1.0003$. 

In general, one needs to consider the finite extension of the shower
at a given time $t'$, expressed via a weight function $w(r^{1},r^{2},h)$,
where $r^{1}$ and $r^{2}$ represent the transverse distance from
the shower axis, and $h$ the longitudinal distance from the shower
front, the latter one moving by definition with the vacuum velocity
of light. Positive $h$ means a position behind the shower front,
and therefore $w$ is non-vanishing only for positive $h$. The weight
will fall off rapidly with increasing distance $r=\sqrt{(r^{1})^{2}+(r^{2})^{2}}$
from the axis. The precise form of $w$ will be discussed in a later
chapter. In principle one needs to convolute the weight $w$ with
the currents, and then compute the potential and field. Due to a translation
invariance (being correct to a very good approximation, since the
index of refraction varies slowly), this is the same as computing
first the potential in PL approximation, and then performing a convolution
as \begin{equation}
A^{\beta}(t,\vec{x})=\int\!\! d^{2}r\!\!\int\!\! dh\, w(\vec{r},h)A_{\mathrm{PL}}^{\beta}(t,x^{\Vert}-h,\vec{x}^{\perp}+\vec{r}).\end{equation}
where $x^{\Vert}$ and $\vec{x}^{\bot}=(x^{\bot1},x^{\bot2})$ are
coordinates parallel and transverse to the shower axis. Defining $\vec{y}^{\bot}=\vec{x}^{\bot}+\vec{r}$,
we get\begin{equation}
A^{\beta}(t,\vec{x})=\int\!\! d^{2}y^{\bot}\!\!\int\!\! dh\, w(\vec{y}^{\bot}-\vec{x}_{\perp},h)A_{\mathrm{PL}}^{\beta}(t,x^{\Vert}-h,\vec{y}^{\bot}).\label{eq:abas}\end{equation}
The electric field is then obtained from the derivatives of $A$. 

\begin{figure}[tbh]
\begin{centering}
\includegraphics[scale=0.4]{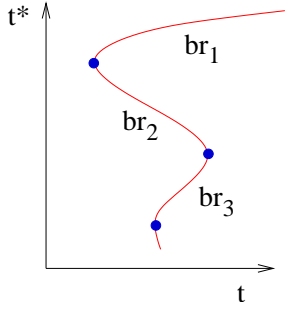}
\par\end{centering}

\vspace{-0.3cm}

\caption{Several branches of the function of $t^{*}$ versus $t$, where $t^{*}$
is the retarded time corresponding to an observer time $t$ . \label{cap:branches}}

\end{figure}

One cannot simply exchange derivation and integration, due to the
presence of singularities as discussed before, and therefore a naive
convolution of $w$ with $\vec{E}_{\mathrm{PL}}$ is not possible:
one needs a more sophisticated treatment of the singularities. So
let us consider the most general case of a multi-valued function $t^{*}$
as a function of the observation time $t$, as sketched in fig. \ref{cap:branches}.
The function is composed of several branches, $\mathrm{br}_{n}$,
limited by certain times $t_{k}$. The derivative $dt^{*}/dt$ becomes
infinite at these branch endpoints, and the point-like potential becomes
singular. This is why we refer to the $t_{k}$ as ''critical times''.
Close to these singularities, we have \begin{equation}
t^{*}-t^{*}(t_{k})\sim|t-t_{k}|^{1/2},\quad\mathrm{and}\quad\frac{dt^{*}}{dt}\sim|t-t_{k}|^{-1/2}.\end{equation}
When evaluating eq. (\ref{eq:abas}), we have to worry about the critical
time for a given observer position $(x^{\Vert}-h,\vec{y}^{\bot})$,
corresponding to the arguments of $A_{\mathrm{PL}}$. In other words,
$t_{k}$ is a function of these variables, i.e. \begin{equation}
t_{k}=t_{k}(x^{\Vert}-h,\vec{y}^{\bot}).\end{equation}
It is useful to define a {}``critical $h$ value'' $h_{k}$, for
given $t$, via\begin{equation}
t=t_{k}(x^{\Vert}-h_{k},\vec{y}^{\bot}),\end{equation}
which allows us to write eq. (\ref{eq:abas}) for a single branch
as\begin{align}
A^{\beta}(t,\vec{x}) & =\int\!\! d^{2}y^{\bot}\!\!\int_{0}^{h_{k}}\!\! dh\, w(\vec{y}^{\bot}-\vec{x}^{\bot},h)\nonumber \\
 & \qquad A_{\mathrm{PL}}^{\beta}(t,x^{\Vert}-h,\vec{y}^{\bot}).\end{align}
Using the integration variable $\lambda=h_{k}-h$, we obtain our master
formula for the vector potential,\begin{align}
A^{\beta}(t,\vec{x}) & =\int\!\! d^{2}y^{\bot}\!\!\int_{0}^{h_{k}}\!\! d\lambda\, w(\vec{y}^{\bot}-\vec{x}^{\bot},h_{k}-\lambda)\nonumber \\
 & \qquad A_{\mathrm{PL}}^{\beta}(t,x^{\Vert}-h_{k}+\lambda,\vec{y}^{\bot}),\end{align}
which is useful because $A_{\mathrm{PL}}$ has a singularity in $\lambda\:$
for\\
$\lambda\to0$, which does not interfere with the derivatives which
have to be performed in order to get the fields. In the following
we keep in mind that $A_{\mathrm{PL}}^{\beta}$ has the following
arguments: the time $t$, the longitudinal variable $x^{\Vert}-h_{k}+\lambda$,
and the transverse variable $\vec{y}_{\perp}$; $w$ has the arguments
$h_{k}-\lambda$ and $\vec{y}_{\perp}-\vec{x}_{\perp}$. We do not
write these arguments explicitly, to simplify the notation. We also
omit the arguments $t,\vec{x}$ of the vector potential. So we write\begin{equation}
A^{\beta}=\int\!\! d^{2}y^{\bot}\!\!\int_{0}^{h_{k}}\!\! d\lambda\, w\, A_{\mathrm{PL}}^{\beta}.\end{equation}
The components of the electric field are \begin{equation}
E^{\Vert}=c(-\frac{\partial A^{0}}{\partial x^{\Vert}}-\frac{\partial A^{\Vert}}{\partial\, ct})\end{equation}
\begin{equation}
E^{\bot i}=c(-\frac{\partial A^{0}}{\partial x^{\bot i}}-\frac{\partial A^{\bot i}}{\partial\, ct}).\end{equation}
Using $A_{\mathrm{PL}}^{i}=\frac{\mu_{0}}{4\pi}\, J^{i}\,|\widetilde{R}V|^{-1}$
and eqs. (\ref{eq:app1},\ref{eq:app2}), the time derivative of the
vector potential may be written as\begin{equation}
\frac{\partial A^{i}}{\partial\, ct}=\int d^{2}y_{\perp}\!\!\int_{0}^{h_{k}}\!\! d\lambda\,\left\{ -w'\, A_{\mathrm{PL}}^{i}+w\,\dot{A}_{\mathrm{PL}}^{i}\right\} ,\end{equation}
with $w'=\partial w/\partial h$, $\dot{A}_{\mathrm{PL}}^{i}=\frac{\mu_{0}}{4\pi}\, K^{i}\,|\widetilde{R}V|^{-1}$,
$K=dJ/dt'$. In principle there is an additional term from the time
derivative of the upper limit of integration, but this contribution
vanishes due to $w(0)=0$ (see next chapter). Concerning the space
derivative, we first compute the derivative with respect to the longitudinal
variable. We find\begin{equation}
-\frac{\partial}{\partial x^{\Vert}}A^{0}=-\int d^{2}y_{\perp}\!\!\int_{0}^{h_{k}}\! w'\, A_{\mathrm{PL}}^{0}\! d\lambda\,\end{equation}
since the total longitudinal space derivative of $A_{\mathrm{PL}}^{0}$
vanishes exactly. The transverse derivatives of the scalar potential
can be expressed in terms of the derivatives $w^{i}=\partial w/\partial r^{i}$
of the weight function $w$ as \begin{equation}
-\frac{\partial}{\partial x^{\bot i}}A^{0}\!=\!\int\! d^{2}y_{\perp}\!\int_{0}^{h_{k}}\!\! d\lambda\, w^{i}\, A_{\mathrm{PL}}^{0}.\end{equation}
The above results for the partial derivatives of the vector potential
$A^{\mu}$ allow us to obtain corresponding expressions for the electric
field. The longitudinal electric field $c(\partial^{\Vert}A^{0}-\partial^{0}A^{\Vert})$
is given as\begin{equation}
E^{\Vert}=-c\!\int\! d^{2}y_{\perp}\!\int_{0}^{h_{k}}\!\! d\lambda\,\left\{ w'\, A_{\mathrm{PL}}^{0}-w'\, A_{\mathrm{PL}}^{\Vert}+w\,\dot{A}_{\mathrm{PL}}^{\Vert}\right\} .\label{eq:ee1}\end{equation}
The transverse field $c(\partial^{\bot i}A^{0}-\partial^{0}A^{\bot i})$
can be written as\begin{equation}
E^{\bot i}=c\int\!\! d^{2}y_{\perp}\!\!\int_{0}^{h_{k}}\!\! d\lambda\left\{ w^{i}\, A_{\mathrm{PL}}^{0}+w'\, A_{\mathrm{PL}}^{\bot}-w\,\dot{A}_{\mathrm{PL}}^{\bot}\right\} .\label{eq:ee2}\end{equation}
The formulas simplify considerably far from the singularity as well
as at the singularity, but we keep the exact expressions, in order
to interpolate correctly between the two extremes. It should be noted
that the above expression concerns a single branch, the complete field
is the sum over all branches.

In the present work we have derived the electric field directly from
the Liénard-Wiechert potentials in the Lorentz gauge without further
approximations. The distribution of the particles in the shower front
over a finite volume is the reason that our final result is not plagued
with singularities in the vicinity of Cherenkov emission. We thus
explicitly include both the near- and the far-field components of
the radiation. In this sense it differs considerably from the calculations
presented in~\citet{alvarez} where an ad-hoc frequency cut-off is
introduced in the calculations of $300$~MHz, and the near-field
component of the electric field is neglected (the Fraunhofer condition).
As can be seen from fig. \ref{fig:anita} below, the data show a considerable
intensity above $300$~MHz. A question that arises in this respect
is the validity of the Fraunhofer condition when Cherenkov effects
come into play which implicitly is assumed in~\citet{alvarez}. Often
the Fraunhofer condition is formulated as $a^{2}\sin^{2}\theta/R<\lambda/2\pi\;,$where
$a$ is the length of the emission trajectory, $\theta$ is the opening
angle from the emission point to the observer, $\lambda$ the wavelength
of the emitted signal, and $R$ the distance from the emission point
to the observer. If there is a single point on $a$ where the Cherenkov
condition is fulfilled, the electric field will diverge at this point
whereas the field is finite at all other points. This implies that
thus the Fraunhofer condition is not valid. A Lorentz-invariant formulation
of the Fraunhofer condition is $a^{2}\sin^{2}\theta/\tilde{R}V<\lambda/2\pi,$
where the distance $R$ is replaced by the retarded distance $\tilde{R}V$.
Since the retarded distance vanishes at the Cherenkov angle this clearly
shows that at this point the Fraunhofer condition is no longer valid
for which reason we have not made this assumption in our approach.

\section{Monte Carlo simulations and fitting procedures: EVA 1.0}

The numerical evaluation of the eqs. (\ref{eq:ee1},\ref{eq:ee2})
is done employing the EVA 1.0 package, which

\begin{figure}[tbh]
\begin{centering}
\includegraphics[scale=0.35]{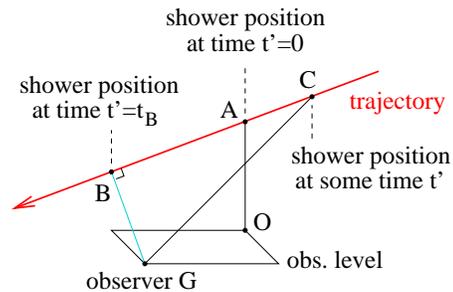}
\par\end{centering}

\caption{Air shower as seen by an observer $G$. The point $B$ is the point
of closest approach with respect to the observer $G$. The point $C$
is the shower position at some time $t'$. The point $B$ corresponds
to the shower position at $t'=t_{B}$ (which may be taken to be zero).
\label{cap:impact}}

\end{figure}

\begin{itemize}
\item provides the weights $w$,
\item provides the currents $J$ needed to compute the potentials $A_{\mathrm{PL}}^{\mu}$,
\item does the numerical integration of eqs. (\ref{eq:ee1},\ref{eq:ee2})
and the summation over branches.
\end{itemize}
Both the weights $w$ and the point-like currents $J$ are obtained
from realistic Monte Carlo simulations of air showers. The EVA package
consists of several elements:

\begin{itemize}
\item the air shower simulation code CX-MC-GEO, including analysis tools
to extract four-currents and the shape of the shower,
\item the automatic fitting procedure FITMC which allows to obtain analytical
expressions for the currents,
\item the EVA program which solves the non-trivial problem to compute the
retarded time and {}``the denominator'' $\tilde{R}V$, for a realistic
index of refraction.
\end{itemize}
We first discuss air showers. They are considered point-like for the
moment, as seen by a far-away observer. The shower is a moving point,
defining a straight line trajectory, see fig. \ref{cap:impact} .%
One defines an {}``observer level'' which is a plane of given altitude
$z$ with respect to the sea level. One defines some arbitrary point
$A$ on the trajectory. The corresponding projection to the observer
level is named $O$ (origin) and the observer position is given in
terms of coordinates $(x,y)$ with respect to $O$. The $x-$axis
is the intersection of the {}``shower plane'' $OAC$ and the observer
level. The angle between the shower trajectory and the vertical axis
$OA$ is referred to as inclination and denoted as $\theta$. In many
applications, $A$ and $O$ coincide: in this case they represent
the impact point. For horizontal showers the two points are different.
The geomagnetic field is specified by an angle $\alpha$ with respect
to the vertical, and an angle $\psi$ with respect to the shower plane
($\psi=0$ means that $\vec{B}$ points towards the shower origin).
One can of course see it the other way round (maybe even more natural):
for a given orientation of the geomagnetic field, $\psi$ defines
the orientation of the shower axis. 

In the EVA framework, one has to specify the altitude $z$, the distance
$a=|OA|$, the inclination $\theta$, the energy $E$ of the shower,
and the observer coordinates $x$, $y$. And in addition the angles
$\alpha$ and $\psi$ and the magnitude $B$ of the geomagnetic field.

The actual air shower simulations are done with a simulation program
called CX-MC-GEO, being part of the EVA package. It is based on CONEX
\citet{conex1,conex2}, which has been developed to do air shower
calculations based on a hybrid technique, which combines Monte Carlo
simulations and numerical solutions of cascade equations. It is also
possible to run CONEX in a pure cascade mode, and this is precisely
what we use. This provides full Monte Carlo air shower simulations,
using EGS4 \citet{egs4} for the electromagnetic cascade, and the
usual hadronic interaction models (QGSJET, EPOS, etc) to simulate
hadronic interactions. 

Two features have been added to CONEX. First of all a magnetic field,
which amounts to replacing the straight line trajectories of charged
particles by curved ones. This concerns both the electromagnetic cascade
and the hadronic one. In addition, analysis tools have been developed,
which allow to get a complete information of charged particle flow
in space and time. These features have already been developed to compute
currents in \citet{klaus1}, so in particular more details about the
implementation of the magnetic field can be found there (though we
did not use the names EVA and CX-MC-GEO yet). We also discuss in \citet{klaus1}
some details about the different internal coordinate systems needed
to extract information about currents. The results shown in \citet{krijn}
were also based on CX-MC-GEO simulations, referred to as CONEX-MC-GEO
at the time.

In \citet{klaus1}, we provide several results concerning particle
numbers and currents for different orientations of the axis with respect
to the geomagnetic field. All the results are still valid, the corresponding
programs did not change since. 

An important ingredient of our approach is the parametrization of
the results (currents $J$, distributions $w$), which have been obtained
from simulations in the form of discrete tables. This is necessary
partly to perform semi-analytical calculations such that numerically
stable functions have to be dealt with without having huge cancellations
in the results. It is especially important for the stable calculation
of Cherenkov effects. It allows for the calculation of a smooth shower
evolution, whereas when working with histogramed distributions in
position and time, it is not possible to reconstruct a continuous
shower evolution and the artificially introduced sudden changes in
the particle trajectories may give rise to spurious radio signals. 

The parametrization of Monte Carlo distributions is done in FITMC.
This program takes the distributions (for currents) as obtained from
the simulations in the form of histograms, to obtain analytical expressions,
using standard minimization procedures. FITMC creates actually computer
code to represent the analytical functions, and this code is then
executed at a later stage. The {}``basic distribution'' is the so-called
{}``electron number $N$'' (which counts the number of electrons
and positrons) as a function of the shower time $t'$, which is fitted
as \begin{align}
N(t') & =A\,\exp\big(B+C(t'+D)\\
 & \qquad\qquad\qquad+E(t'+F)^{2}+G(t'+H)^{3}\,\big).\nonumber \end{align}

As an illustration, we show here the case of a shower with an initial
energy of $5\cdot10^{17}$eV, an inclination $\theta=27^{0}$, and
an azimuth angle $\psi=0^{0}$, defined with respect to the magnetic
north pole. The angle $\psi$ refers to the origin of the shower.
So $\psi=0^{0}$ thus implies that the shower moves from north to
south.  We consider the magnetic field at the {\small CODALEMA} site,
i.e. $|\vec{B}|=47.3\mu T$ and $\alpha=153^{0}$, so the shower makes
an angle of roughly $54^{0}$ with the magnetic field. 

\begin{figure}[tbh]
\begin{centering}
\includegraphics[scale=0.23,angle=270]{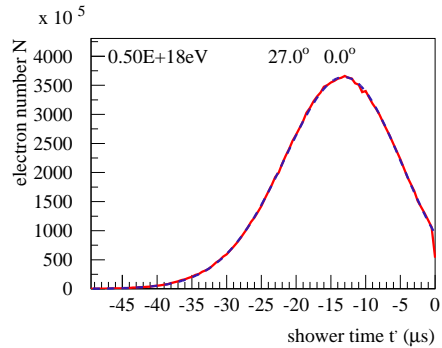}
\par\end{centering}

\caption{The number $N$ of electrons and positrons, as a function of the shower
time $t'$ for a shower with an inclination $\theta=27^{0}$ and an
azimuth angle $\psi=0^{0}$ with respect to the magnetic north pole.
The full red line represents the simulation result, the dashed blue
line is the fit.\label{cap:nel00}}

\end{figure}

In fig.\ref{cap:nel00}, %
we plot the electron number $N$, as a function of the shower time
$t'$, for a simulated single event, together with the fit curve.
A thinning procedure has been applied (here and in the following)
to obtain the shown simulation results. The time $t'=0$ refers to
the point of closest approach with respect to an observer at $x=0$,
\textbf{$y=500\,$}m, $z=140\,$m. We suppose $a=0$ (so the shower
hits the ground at $x=y=0$). 

\begin{figure}[tbh]
\begin{centering}
\includegraphics[scale=0.23,angle=270]{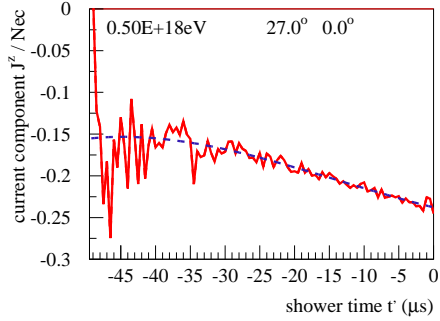}
\par\end{centering}

\caption{The longitudinal current component $J^{z}$, divided by $Nec$, as
a function of the shower time $t'$. The full red line represents
the simulation result, the dashed blue line is the fit.\label{cap:jz}}

\end{figure}

\begin{figure}[b]
\begin{centering}
\includegraphics[scale=0.23,angle=270]{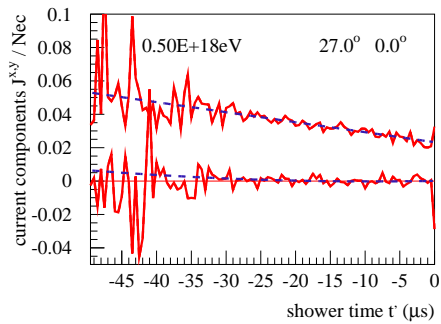}
\par\end{centering}

\caption{The transverse current components $J^{x}$ (lower lines) and $J^{y}$
(upper lines), divided by $Nec$, as a function of the shower time
$t'$. The full red lines represents the simulation result, the dashed
blue lines are the fits.\label{cap:jxy}}

\end{figure}

The magnitudes of the components $J^{\mu}$ of the currents have a
similar $t'$ dependence as $N(t')$. Therefore we parametrize the
ratios $J^{\mu}/(Nec)$, with $N$ being the electron number, $e$
the elementary charge, and $c$ the velocity of light. We use the
following parametric form:\begin{equation}
\frac{J^{\mu}(t')}{N(t')ec}=A+B(x+C)+D(x+E)^{2}+F(x+G)^{3}.\end{equation}
In fig. \ref{cap:jz}, %
we plot the longitudinal current component $J^{z}$ (divided by $Nec)$,
as a function of the shower time $t'$, for a simulated single event,
together with the fit curve. At early times - far away from the shower
maximum -- there are of course large statistical fluctuations. But
since $N(t')$ is very small here, this region does not contribute
to the pulse. In fig. \ref{cap:jxy}, %
we plot the transverse current components $J^{x}$ and $J^{y}$, (divided
by $Nec)$, as a function of the shower time $t'$, for a simulated
single event, together with the fit curves.

The EVA program uses these analytical fit functions for the current
components, \begin{equation}
J^{\mu}=\left\{ \frac{J^{\mu}}{Nec}\right\} _{fit}\cdot N_{fit}\cdot e\cdot c,\end{equation}
to compute the vector potential. The currents have to be evaluated
at $t'=t^{*}$, the retarded time. The central part of EVA is actually
the determination of the retarded time $t^{*}(t,\vec{x}$) for a given
observer position. This is quite involved -- in case of a realistic
index of refraction -- and described in detail in \citet{klaus1}
(again without referring to EVA, but these are exactly the same programs
being used). A results of such a calculation is shown in fig. \ref{fig:trett}.

A new feature compared to \citet{klaus1} -- and most relevant for
this paper -- is the possibility to obtain information about the shape
of the shower via the weight function $w$. The weight function $w$
is not perfectly cylindrically symmetric, due to the geo-magnetic
field but also due to statistical fluctuations, since we are considering
individual Monte Carlo events. However, in this paper we will neglect
these tiny deviation from symmetry, and consider a weight function
$w(r,h)$ depending only on the two variables $r$ and $h$, related
to the general weight function as \begin{equation}
w(r,h)=2\pi r\, w(\vec{r},h).\end{equation}
The lateral coordinate $r$ measures the transverse distance from
the shower axis, the longitudinal coordinate $h$ is meant to be the
distance behind the shower front. This front is a hypothetical plane
moving parallel to the shower axis with the velocity of light $c$,
such that all the particles are behind this front, expressed by a
positive value of $h$. We will express the weight function as\begin{equation}
w(r,h)=w_{1}(r)\, w_{2}(r,h),\end{equation}
with $\int dr\, w_{1}(r)=1$, and with $\int dh\, w_{2}(r,h)=1$ for
all values of $r$. 

We use again CX-MC-GEO to obtain $w$, then FITMC to obtain an analytical
function, which is later used in the EVA program to compute the fields,
based on the formulas described in the preceding chapter. All the
simulation results shown in the following are based on the the same
shower, mentioned earlier when discussing currents.

We first investigate the radial distribution $w_{1}(r)$.%
\begin{figure}[tbh]
\begin{centering}
\includegraphics[scale=0.23,angle=270]{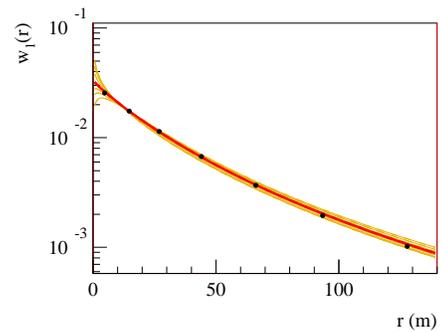}
\par\end{centering}

\caption{The radial distribution $w_{1}(r)$. The thin yellow lines correspond
to different times, the points represent an average, and the thick
red line corresponds to a fit (see text). \label{cap:The-radial-distribution}}

\end{figure}
In fig. \ref{cap:The-radial-distribution}, we show the radial distribution
as obtained from the Monte Carlo simulation. The thin lines correspond
to different times $t'$, between $-25\,\mu s$ and $-5\,\mu s$.
The points represent an average over all times, and also averaged
over $r$--bins. Since the time dependence is quite small, we will
use the radial distribution at the shower maximum $t'_{\max}$ as
time-independent distribution $w_{1}(r)$. The thick red line corresponds
to a fit to the Monte Carlo data, using the form\begin{equation}
w_{1}(r)=\frac{\Gamma(4.5-s)}{\Gamma(s)\Gamma(4.5-2s)}\left(\frac{r}{r_{0}}\right)^{s-1}\left(\frac{r}{r_{0}}+1\right)^{s-4.5},\end{equation}
with fit parameters $r_{0}$ and $s$ (providing an excellent fit).

\begin{figure}[tbh]
\begin{centering}
\includegraphics[scale=0.23,angle=270]{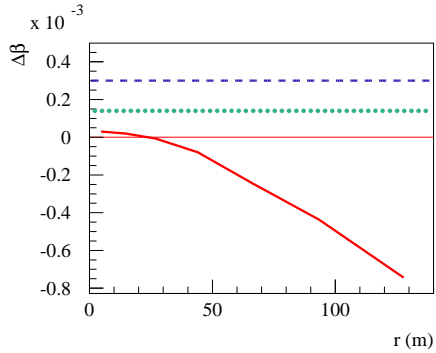}
\par\end{centering}

\caption{The longitudinal velocity difference $\Delta\beta$ versus $r$. We
show the results for realistic simulations (thick red solid line)
and for $\gamma=60$ (green dotted line). Also shown: the value $1-1/n_{\mathrm{ground}}$(blue
dashed line). \label{cap:The-longitudinal-velocity}}

\end{figure}

Knowing $w_{1}(r)$, we now investigate how far the particles are
moving behind the shower front, expressed in terms of the longitudinal
distance $h$, for a given transverse distance $r$. From the above
simulation, we obtain easily the mean distance $\bar{h}$ at a given
$r$. We find a perfectly linear time dependence, of the form\begin{equation}
\bar{h}=h_{\mathrm{front}}+c\Delta\beta\, t',\end{equation}
where $\Delta\beta$ can be obtained from fitting time dependence
at different distances $r$, the result is shown in fig. \ref{cap:The-longitudinal-velocity}
as solid line. %
The quantity $\Delta\beta$ represents the velocity difference (in
units of $c$) with respect to the the shower front, which itself
moves with velocity $c$. So the velocity of the {}``average position''
of the shower is $1-\Delta\beta$. Also shown in fig. \ref{cap:The-longitudinal-velocity},
as dashed line, is the value $1-1/n_{\mathrm{ground}}$, corresponding
to the velocity of light in air with $n_{\mathrm{ground}}=1.0003$.
And we also plot as dotted line the $\Delta\beta$ obtained from $\gamma=60$,
corresponding to the average electron energy. The simulated curve
(thick full line) is considerably below this dashed and the dotted
curves, which means that the velocity of the average positions is
larger than $c/n_{\mathrm{ground}}$, it is also larger than the velocity
of the average electron. The simulated velocity is even (slightly)
larger than $c$. This is due to the fact that matter is moving on
the average from inside (small $r$) to outside (large $r$), and
the average $\bar{h}$ decreases with decreasing distance $r$. But
the effect is small, the deviation of the shower velocity from $c$
is less than 1/1000. %
\begin{figure}[tbh]
\begin{centering}
\includegraphics[scale=0.23,angle=270]{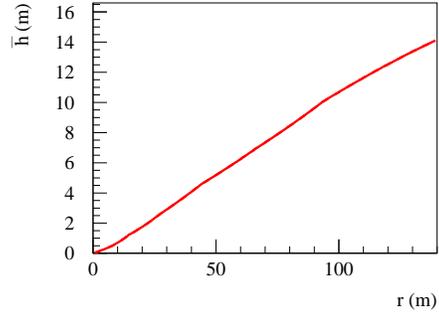}
\par\end{centering}

\caption{The mean value $\bar{h}$ for given values of the lateral distance
$r$.\label{cap:The-mean-value}}

\end{figure}
\begin{figure}[tbh]
\begin{centering}
\includegraphics[scale=0.23,angle=270]{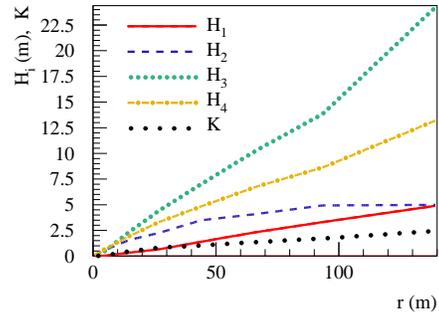}
\par\end{centering}

\caption{The parameters $H_{1}$, $H_{2}$, $H_{3}$, $H_{4}$, and $K$ as
a function of the lateral distance: $H_{1}$ (full line) , $H_{2}$
(dashed line) , $H_{3}$ (dotted line), $H_{4}$ (dashed-dotted line)
, $K$ (wide-dotted line) , \label{cap:The-parameters-}}

\end{figure}
\begin{figure}[tbh]
\begin{centering}
\includegraphics[scale=0.23,angle=270]{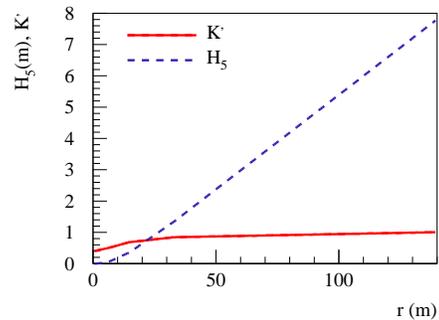}
\par\end{centering}

\caption{The parameters $H_{5}$ and $K'$ as a function of the lateral distance:
$K'$ (full line) , $H_{5}$ (dashed line) \label{cap:The-parameters-2}}

\end{figure}
We will ignore the small time dependence for the moment, and consider
in the following quantities at $t_{\max}$. To get some idea about
the typical scales of the $h$--distribution $w_{2}(r,h)$, for a
given value of $r$, we determine the mean value $\bar{h}$, as shown
in fig. \ref{cap:The-mean-value}. The mean value $\bar{h}$ is almost
a linear function of the distance $r$, and for $r=100\,\mathrm{m}$
we get an average $h$ of roughly $10\,\mathrm{m}$.

\begin{figure}[tbh]
\begin{centering}
~
\par\end{centering}

\begin{centering}
\includegraphics[scale=0.23,angle=270]{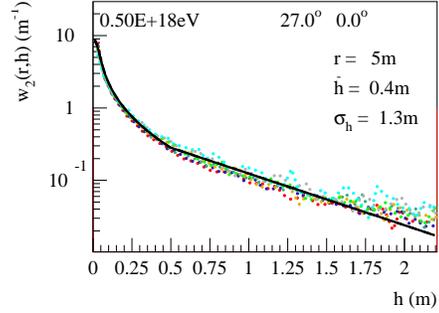}
\par\end{centering}

\caption{The distribution $w_{2}(r,h)$ for $r=5\,\mathrm{m}$. The full black
line represents the fit, the dotted lines are simulation results for
different times. \label{cap:w2a}}

\end{figure}

\begin{figure}[tbh]
\begin{centering}
~
\par\end{centering}

\begin{centering}
\includegraphics[scale=0.23,angle=270]{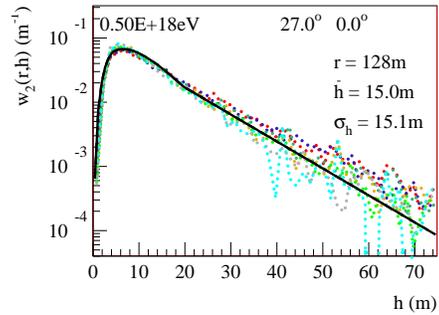}
\par\end{centering}

\caption{The distribution $w_{2}(r,h)$ for $r=128\,\mathrm{m}$. The full
black line represents the fit, the dotted lines are simulation results
for different times. \label{cap:w2b}}

\end{figure}

\begin{figure}[tbh]
\begin{centering}
~
\par\end{centering}

\begin{centering}
\includegraphics[scale=0.23,angle=270]{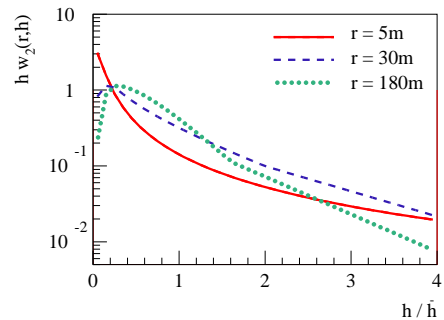}
\par\end{centering}

\caption{The distribution $w_{2}(r,h)$ for $r=5\,\mathrm{m}$(full line),
$r=30\,\mathrm{m}$ (dashed line), and $r=180\,\mathrm{m}$ (dotted
line). \label{cap:w2c}}

\end{figure}

The $w_{2}$ distribution is obtained by fitting Monte Carlo data
in a range $h$ between zero and $5\,\bar{h}$, for given $r$. We
use\begin{equation}
w_{2}(r,h)=\left\{ \begin{array}{ccc}
w_{2}^{\mathrm{MGD}}(r,h) & \mathrm{for} & r>r_{0}\\
w_{2}^{\mathrm{IGD}}(r,h) & \mathrm{for} & r\leq r_{0}\end{array},\right.\end{equation}
with $w_{2}^{\mathrm{MGD}}$ being a {}``modified gamma distribution''
of the form\begin{equation}
w_{2}^{\mathrm{MGD}}(r,h)=\frac{H(r,h)\, G(r,h)}{N(r)},\end{equation}
with\begin{equation}
H(r,h)=\Theta(H_{1}-h)\left(2\left(\frac{h}{H_{1}}\right)-\left(\frac{h}{H_{1}}\right)^{2}\right)+\Theta(h-H_{1}),\end{equation}
and\begin{eqnarray}
G(r,h) & = & \Theta(H_{3}-h)\left(h^{K-1}\, e^{-h/H_{2}}\right),\\
 &  & +\Theta(h-H_{3})\left(H_{3}{}^{K-1}\, e^{-H_{3}/H_{2}}e^{-(h-H_{3})/H_{4}}\right),\nonumber \end{eqnarray}
with $N$ being a normalization constant such that $\int dh\, w_{2}(r,h)=1$.
The function $w_{2}^{\mathrm{IGD}}$ is an {}``inverse gamma distribution''
of the form\[
w_{2}^{\mathrm{IGD}}(r,h)=\frac{(H_{5})^{K'}}{\Gamma(K')}h^{-K'-1}e^{-H_{5}/h}.\]
We use $r_{0}=20\mathrm{m}$. The $r$--dependence is hidden in the
parameters $H_{1}$, $H_{2}$, $H_{3}$, $H_{4}$, $H_{5}$, $K$,
and $K'$. In figs. \ref{cap:The-parameters-} and \ref{cap:The-parameters-2},
we plot the parameters, as obtained from fitting the Monte Carlo data.
All parameters grow with increasing distance $r$. Whereas $H_{2}$
seems to saturate, all the other parameters grow roughly linearly
with $r$. With these parameters, we get good fits for $h$ values
up to five times the mean. In figs. \ref{cap:w2a} and \ref{cap:w2b},
we show the fits of $w_{2}$ together with Monte Carlo simulation
results for different times. In fig. \ref{cap:w2c}, we show the fitted
$w_{2}$ curves for three different values of $r$, conveniently plotted
as $h\, w_{2}$ versus $h/\bar{h}$, where one clearly sees the evolution
of the shape with $r$.%

The reason to switch between $w_{2}^{\mathrm{MGD}}$ and $w_{2}^{\mathrm{IGD}}$
at $r_{0}=20$~m becomes clear from figures \ref{cap:w2a} through
\ref{cap:w2c}. From figure \ref{cap:w2c} it can be seen clearly
that the particle distribution as obtained from the Monte-Carlo simulations
behaves quite differently close to the shower axis as compared to
the distribution at large distances. This different behavior requires
the use of different fit-functions in both regimes. At large distances
from the core, the parametrization of $w_{2}^{\mathrm{MGD}}$ reproduces
the MC result accurately. At small distances, it is important to have
a smooth parametrization without jumps in the first derivatives, which
is the case when using $w_{2}^{\mathrm{IGD}}$.

The above fit function $w_{2}^{\mathrm{IGD}}$ leads to a delta-peak
at $r=0$. To still obtain numerical stability, a cut-off for the
values $K'$ and $H_{5}$ is introduced such that the width of $w_{2}^{\mathrm{IGD}}$
is 1~mm. Since most of the particles are located within $r=\delta x^{\perp}=1$~m
from the shower axis, the path difference between signals emitted
at this distance on both sides of the shower axis acts as the important
length scale in this regime. We estimate this path difference $\delta R$
for a constant index of refraction equal to $n=1.0003$ : we have
$\delta{R}\approx\frac{\partial R}{\partial x^{\perp}}\delta x^{\perp}\approx\sqrt{n^{2}\beta^{2}-1}\delta x^{\perp}\approx3$~cm.
Here we use that at the Cherenkov time (critical time for $h=0$),
we have $R^{0}=n\beta x^{||}$, and $x_{c}^{||}=\sqrt{n^{2}\beta^{2}-1}x_{c}^{\perp}$
\citet{arena}. So a cut-off of $w_{2}^{\mathrm{IGD}}\approx1$~mm
should give stable results. This has been tested numerically.

\section{Time signals}

\begin{figure}[tbh]
\hspace*{-0.4cm}\includegraphics[angle=270,scale=0.33]{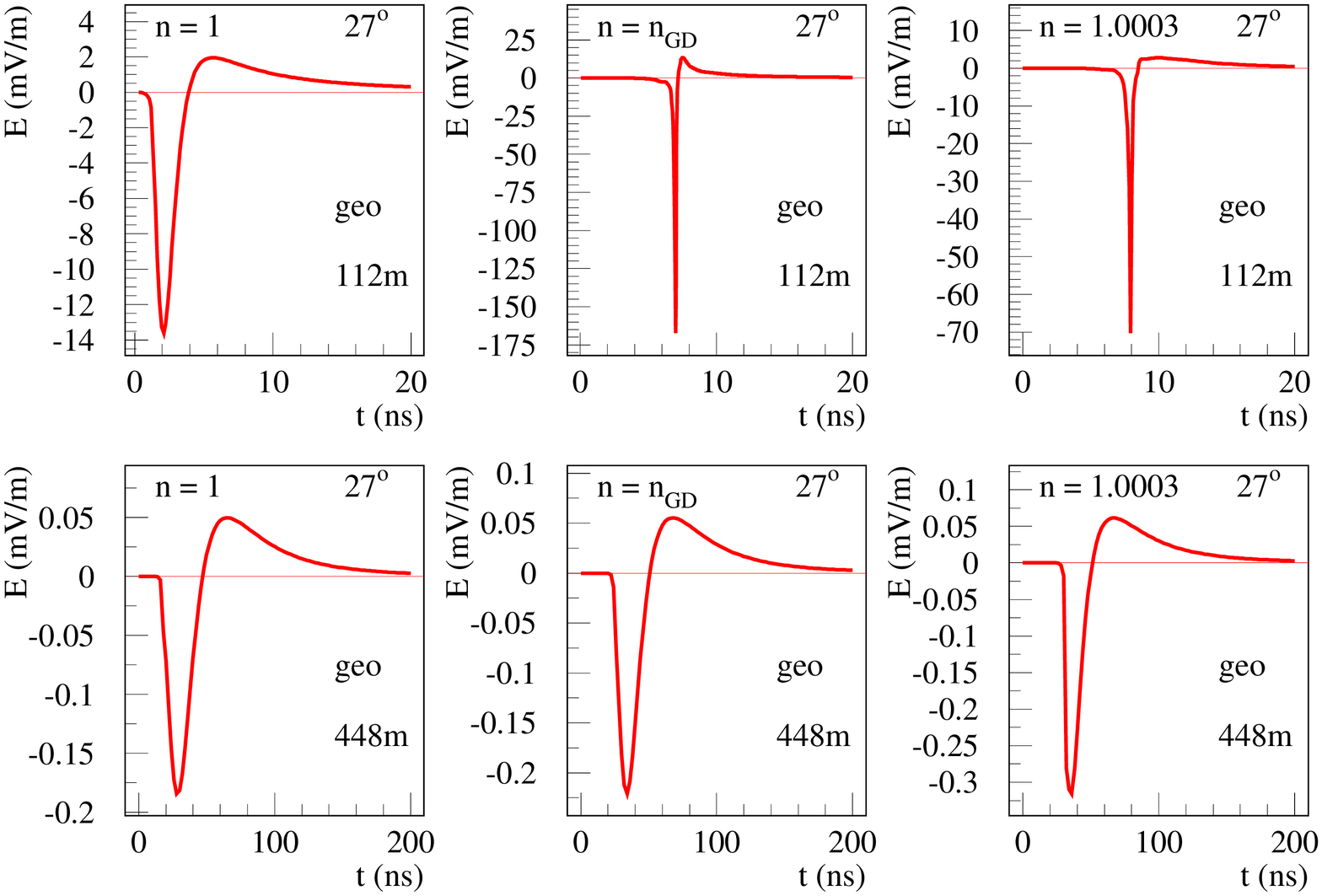}

\caption{The $y$ component of the geomagnetic contribution to the electric
field as a function of the observer time $t$ in ns, for an observer
distance of 112 m (upper panel) and 448 m (lower panel) We compare
different options for the index of refraction $n$, namely $n=1$(left),
$n=n_{\mathrm{GD}}$(middle), and $n=1.0003$(right).\label{fig:e1}}

\end{figure}
\begin{figure}[tbh]
\hspace*{-0.4cm}\includegraphics[angle=270,scale=0.33]{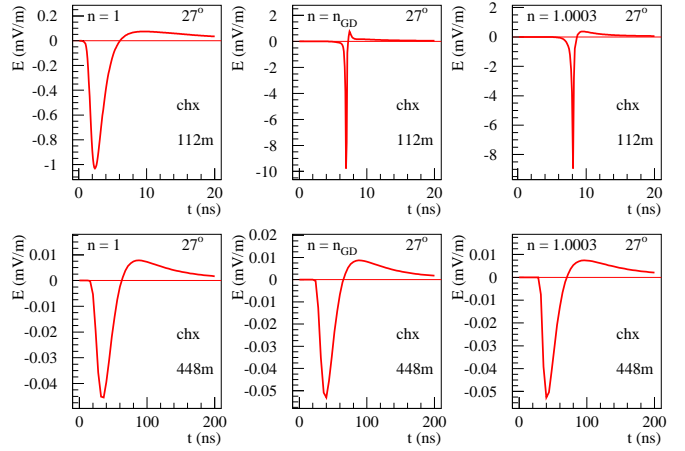}

\caption{The $x$ component of the charge excess contribution to the electric
field as a function of the observer time $t$ in ns, for an observer
distance of 112 m (upper panel) and 448 m (lower panel) We compare
different options for the index of refraction $n$, namely $n=1$(left),
$n=n_{\mathrm{GD}}$(middle), and $n=1.0003$(right).\label{fig:e2}}

\end{figure}
As already said, the eqs. (\ref{eq:ee1},\ref{eq:ee2}) are evaluated
employing the EVA 1.0 package, which provides the weights $w$, the
currents $J$, the denominators $\tilde{R}V$, and the integration
procedures, as discussed in the previous chapter. We first consider
the same {}``reference shower'' (initial energy of $5\cdot10^{17}$eV,
inclination $27^{o}$) already discussed there. We will distinguish
between the geomagnetic contribution (caused by the currents due to
the geomagnetic field) and the contributions due to charge excess.
In figs. \ref{fig:e1} and \ref{fig:e2}, we show the results for
the two contributions, for two different observer positions: 112 and
448 meters to the south of the impact point. We compare the realistic
scenario ($n=n_{\mathrm{GD}})$ with the two {}``limiting cases''
$n=1$ and $n=1.0003$. One can clearly see big differences between
the three scenarios, up to a factor of ten in width and magnitude.
We also see, even in the realistic case ($n=n_{\mathrm{GD}})$, the
appearance of {}``Cherenkov-like'' behavior, with very sharp peaks.%
\begin{figure}[b]
\hspace*{-0.4cm}\includegraphics[angle=270,scale=0.33]{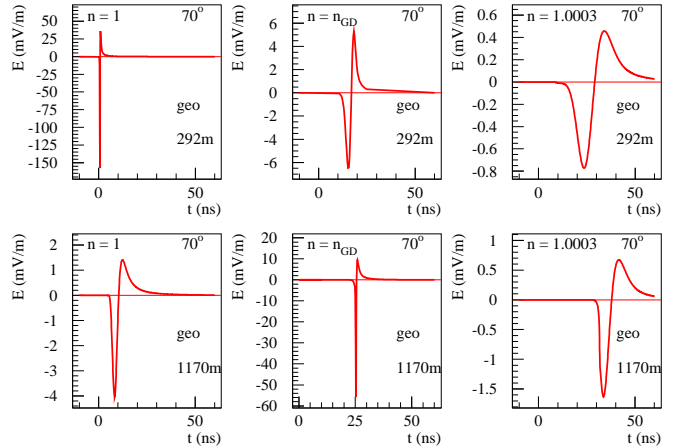}

\caption{Same as fig \ref{fig:e1}, but here we consider a more inclined shower.\label{fig:e3}}

\end{figure}
\begin{figure}[tbh]
\hspace*{-0.4cm}\includegraphics[angle=270,scale=0.33]{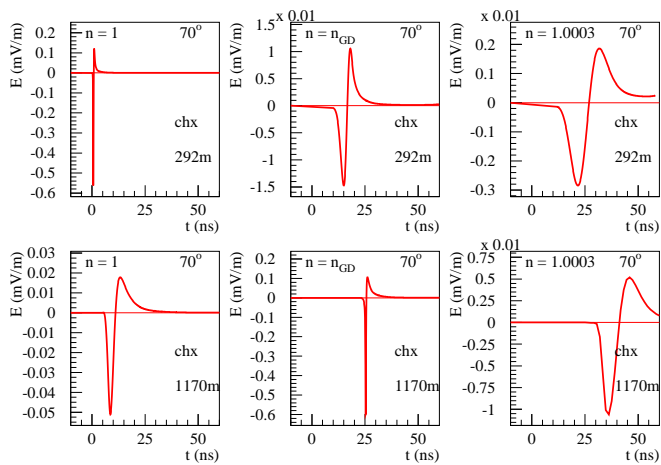}

\caption{Same as fig \ref{fig:e2}, but here we consider a more inclined shower.\label{fig:e4}}

\end{figure}
In figs. \ref{fig:e3} and \ref{fig:e4}, we consider a more inclined
shower ($70^{o}$), for two different observer positions: 292 and
1170 meters to the south of the impact point. The differences between
the realistic case ($n=n_{\mathrm{GD}})$ and the two {}``limiting
cases'' is even bigger: more than a factor of 100 in width and magnitude!

\section{Geomagnetic Cherenkov radiation}

As shown in the last chapter, a realistic treatment of the index of
refraction in the atmosphere seems to be very crucial for the forms
of the electromagnetic pulses. Can this be seen in experiments? What
exactly should one look for?

To answer these questions we will discuss in the following frequency
spectra. As shown in chapter II, the fields are sums of terms of the
form (up to factors) \begin{equation}
\int\, dV\mathrm{\: pancake}\:\times\:\mathrm{currents}\:\times\left(\frac{dt^{*}}{dt}\right),\end{equation}
where {}``currents'' and {}``pancake'' refer to respectively the
pointlike currents and the current distributions in the pancake, or
its derivatives. The quantity $dV$ is a pancake volume element. The
currents and the {}``Cherenkov term'' $dt^{*}/dt$ are taken at
the retarded time $t'=t^{*}$, for a given observer time and position.
Let us consider the evolution of an air shower in time $t'$. The
currents are essentially proportional to the electron number $N_{e}(t')$
of the shower, the so-called {}``profile''. We define $t{}_{P}^{*}$
to be an emission time (retarded time) corresponding to the profile
maximum, also referred to as shower maximum. Another important quantity
is the Cherenkov time $t_{C}^{*}$, corresponding to the time where
$dt^{*}/dt$ becomes singular.

The electric field contains actually terms governed by the derivatives
of the currents, and therefore by the derivative of the profile. We
consider therefore the expression {}``shower maximum'' to represent
the actual maximum of the profile or of its derivative.

\begin{figure}[tbh]
\begin{centering}
\includegraphics[angle=270,scale=0.23]{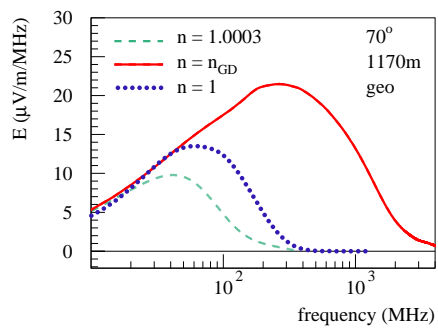}
\par\end{centering}

\caption{Fourier transform of geomagnetic component of the 70 degrees inclined
shower observed at 1170 meters from the shower core. We plot the modulus
of the Fourier transform.\label{fig:fft}}

\end{figure}

A strong signal is expected when the two times $t{}_{P}^{*}$ and
$t_{C}^{*}$ coincide. Such a situation is shown in fig. \ref{fig:fft},
where we plot the Fourier spectrum for the geomagnetic component of
the electromagnetic field for the $70$~degrees inclined shower discussed
in the previous chapter, with an initial energy of $5\cdot10^{17}$eV,
and an observer positioned at a distance of $d=1170$~m to the east
of the impact point of the shower, corresponding to an impact parameter
$b$ of around $400$~m. At this distance the shower maximum occurs
at the Cherenkov time for a realistic index of refraction. The realistic
case ($n=n_{GD}$) contains very high frequency components up to several
GHz as one would expect from the sharp peak in figure 18. The two
limiting cases peak at lower frequencies below $100$~MHz. 

\begin{figure}[tbh]
\begin{centering}
\includegraphics[scale=0.23]{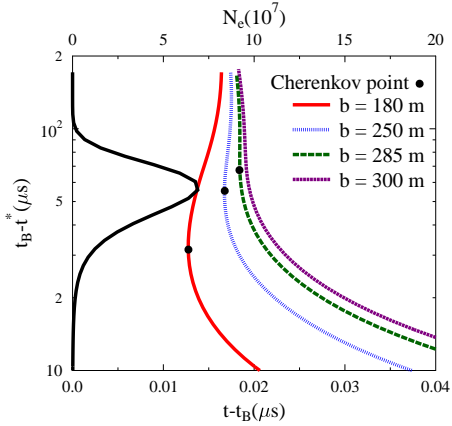}
\par\end{centering}

\caption{The shower profile as a function of $t^{*}$ (black line) and the
retarded times $t^{*}$ as a function of the observer time $t$, relative
to the time of closest approach $t_{B}$ (red, blue, and magenta curves).
The {}``Cherenkov points'' correspond to the Cherenkov times (where
$dt^{*}/dt$ is singular). \label{fig:trett}}

\end{figure}
In the following, we discuss some very interesting features by taking
the example of a $60$~degrees inclined shower with an initial energy
of $10^{17}$\,eV, moving from west to east, in a magnetic field
of strength 24.3$\mu\mathrm{T}$ and an inclination $\alpha$ of $54^{o}$
(Auger site). The observer is positioned to the east of the impact
point. We will use the impact parameter rather than the horizontal
distance (as in the examples before) to characterize the observer
position. 

\begin{figure}[b]
\includegraphics[scale=0.29]{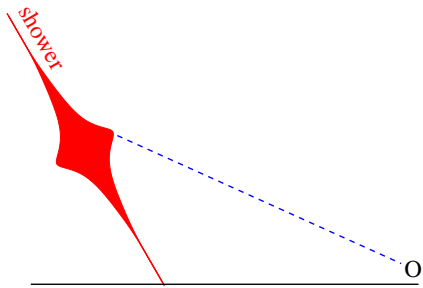}

\caption{The observer $O$ receives {}``normal emission'' from around the
shower maximum.\label{fig:show1}}

\end{figure}

In fig. \ref{fig:trett}, we plot the shower profile $N_{e}$ as a
function of the retarded time $t^{*}$, together the retarded times
$t^{*}$ as a function of the observer time $t$, for three different
choices of the impact parameter. For large values of $b$ (above 285m),
like the case of 300 meters (magenta curve), there is no Cherenkov
time, the function $t^{*}(t)$ is single valued, the derivative is
always finite. We have {}``normal'' emission, coming from around
the the maximum of the profile corresponding to $t'=t_{P}^{*}$, see
fig. \ref{fig:show1}. The form of the time signal is determined by
the profile, we expect maximum frequencies around few hundred MHz,
as confirmed by the calculation shown in fig. \ref{fig:flux}. At
impact parameters smaller than 285 meters, the function $t^{*}(t)$
starts to become double valued, so we we start observing a Cherenkov
time. At 250 meters, the Cherenkov time coincides with the shower
maximum, we have %
\begin{figure}[tbh]
\includegraphics[scale=0.29]{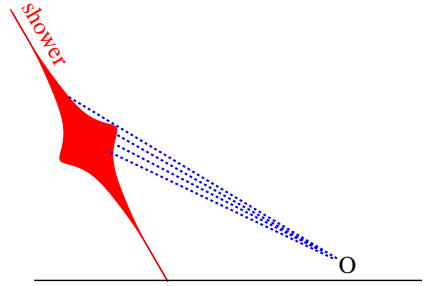}

\caption{The observer $O$ receives {}``Cherenkov emission'' from around
the shower maximum.\label{fig:show2}}

\end{figure}
\begin{figure}[tbh]
\includegraphics[scale=0.29]{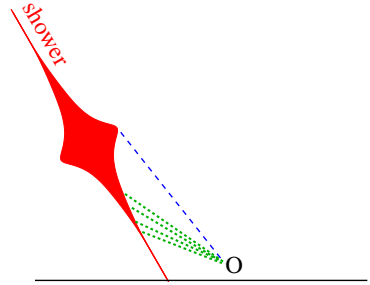}

\caption{The observer $O$ receives both {}``normal emission'' from around
the shower maximum and {}``Cherenkov emission'' from later times.\label{fig:show3}}

\end{figure}
Cherenkov emissions from around the shower maximum. This means that
due to $dt^{*}/dt=\infty$, the emissions from a broad region around
the maximum will be {}``compressed'' and arrive almost at the same
time at the observer, as sketched in fig. \ref{fig:show2}. This leads
to a strong and very short signal. Since the singularity is integrated
over, as explained in chapter II, the actual width of the same signal
is determined by the current distributions in the pancake, and we
expect frequencies around a GHz, as confirmed by the calculation shown
in fig. \ref{fig:flux}.

\begin{figure}[b]
\begin{centering}
\includegraphics[scale=0.23]{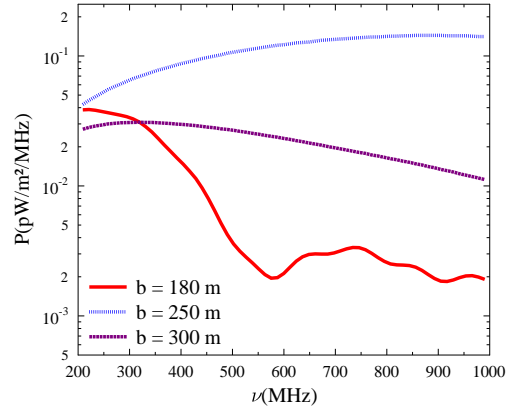}
\par\end{centering}

\caption{Flux densities for radio emission from a $10^{17}$\,eV energy shower
at 60$^{o}$ zenith angle for impact parameters of 180, 250, and 300
meters.\label{fig:flux}}

\end{figure}
 If the observer is even closer to the shower, for example at an impact
parameter of 180 meters, we still have a Cherenkov time, but this
time is now significantly later than the shower maximum time (see
the dot on the red curve in fig. \ref{fig:trett}). Here we may have
a very interesting situation: the observer may receive {}``normal''
emission from around the shower maximum, but at the same time he may
receive a significant contribution from much later, around the Cherenkov
time, which again due to the Cherenkov effect (signal compression)
will be relatively strong and short (high frequency, order of GHz).
This situation is sketched in fig. \ref{fig:show3}. The calculations
in fig. \ref{fig:flux} show (as expected) two distinct peaks, one
at small frequencies due to the normal emission from the shower maximum,
and a second peak at high frequencies due to Cherenkov emission at
much later times.

So our approach predicts not only high frequency components due to
the geomagnetic Cherenkov effect, but in addition a double peak structure
which reflects the simultaneous reception of signals from very different
positions of the shower: {}``normal'' emissions from around the
maximum, and Cherenkov emission from much later times.

\section{Comparing to data}

This geomagnetic Cherenkov radiation might have been observed by the
ANITA-collaboration~\citet{anita}, where pulses have been measured
in the $200$-$1200$~MHz band. Furthermore, since these high frequency
components occur only at the Cherenkov distance, upon applying a high-pass
filter a clear Cherenkov ring should become visible in the LDF. The
radius of this ring contains direct information of the shower maximum
and thus the chemical composition of the original cosmic ray. New
experiments at the Pierre Auger observatory~\citet{auger1,auger2},
and LOFAR~\citet{lofar} should be able to measure the LDF in more
detail, where first hints of \char`\"{}Cherenkov-like\char`\"{} effects
might have been observed~\citet{lopes,lofar2}. 

The key result of our present work is the prediction of a sizable
power emitted at higher frequencies, and a possible double peak structure
with one peak at high frequencies. There exist only few observations
where the spectrum over a large frequency range has been measured.
A good example was published recently by the ANITA collaboration~\citet{Hoover10},
showing the summed power of two cosmic-ray events for the range of
300-900 MHz. 

In this measurement no indication is given of the arrival direction
and the energy of the initiating cosmic ray, only that it most probably
came from a relatively large zenith angle. The azimuth angle is unknown.
Therefore also the air density along the path of the air shower is
unknown, as well as its orientation with respect to the magnetic field.
All this makes any quantitative comparison impossible. To get at least
a qualitative understanding, we compare the data with the result of
a simulation for a cosmic-ray at a zenith angle of $60^{\circ}$ ,
moving from west to east, in a magnetic field corresponding to the
Auger site, with an observer east to the impact point, for various
impact parameters $b$ -- the same situation as discussed in the previous
chapter (changing the energy and the arrival direction of the cosmic
ray will not change the qualitative discussion). 

\begin{figure}[t]

\begin{centering}
\includegraphics[scale=0.25]{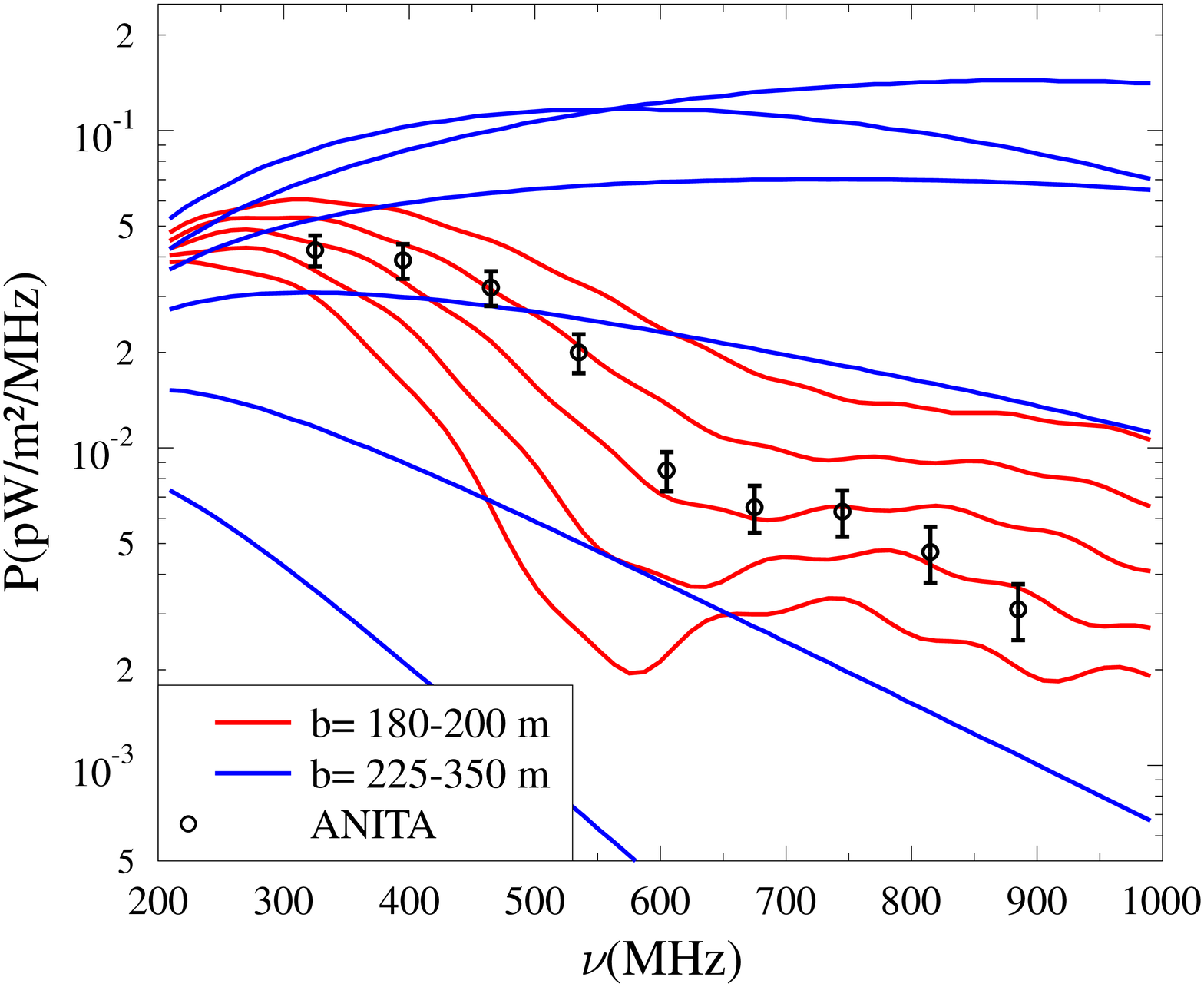}
\par\end{centering}

\caption{The predicted flux densities for radio emission from a $10^{17}$\,eV
energy shower at 60$^{o}$ zenith angle for various impact parameters
$b$ are compared to the data for the sum of two events as measured
by the ANITA balloon mission~\citet{Hoover10}, where the data are
taken from fig. 3 of that publication.\label{fig:anita}}

\end{figure}

In fig. \ref{fig:anita}, %
we compare the data with our simulation results. We show blue curves
corresponding to 350 - 225 m, from bottom to top for the leftmost
value. The red curves refer to 200 - 180m, from top to bottom.

From the discussion of the last chapter, we easily understand the
different theoretical curves: for large impact parameters 350, 325,
300m) we have the situation corresponding to fig. \ref{fig:show1}:
normal emission from around the shower maximum dominates. For impact
parameters around 250 meters, we have Cherenkov emission from around
the shower maximum, as in fig. \ref{fig:show2}, we get strong signals
at large frequencies (GHz). Then finally below 200 meters, we have
the situation sketched in fig. \ref{fig:show3}: a double peak structure
due to simultaneously arriving signals from very different positions
of the shower: {}``normal'' emissions from around the maximum, and
Cherenkov emission from later times.

Although energy and inclination of the measured showers are unknown,
it is nevertheless clear that the data show a structure similar to
the transition region towards a double peak  behavior, as predicted
in our calculations shown in fig. \ref{fig:anita}.

\section{Summary}

We presented a realistic calculation of coherent electro-magnetic
radiation from air showers initiated by ultra-high energy cosmic rays.
The underlying currents are obtained from three-dimensional Monte
Carlo simulations of air showers in a realistic geo-magnetic field.
We take into account the correct shape of the particle distribution
in a shower at a given time. The numerical procedures -- simulations,
fitting procedures, convolutions, referred to as EVA 1.0 -- have been
discussed. We showed the importance of a correct treatment of the
index of refraction in air, given by the law of Gladstone and Dale:
using the correct index of refraction $n_{\mathrm{GD}}$ gives very
different results compared to a simplified treatment using a constant
index, with differences in width and magnitude up to a factor of 100.
The new treatment leads in particular to important emission at high
frequencies (GHz). In certain cases, double peak structures are predicted,
due to signals arriving simultaneously from different positions of
the shower: {}``normal'' emissions from around the maximum, and
Cherenkov emission from later times.

\appendix

\section{\noindent Some derivatives \label{sec:Derivatives-of-the}}

\noindent \textbf{Theorem:} The quantity $h_{k}$ is a function of
$t$ and $x^{\Vert}$(the transverse coordinates are not considered
here). Its derivatives are \begin{equation}
\frac{d}{d\, ct}h_{k}=-1,\quad\frac{d}{dx^{\Vert}}h_{k}=1.\label{eq:app1}\end{equation}
The time derivatives of $ct^{*}$ and $\tilde{R}V$ vanish:\begin{equation}
\frac{d}{d\, ct}ct^{*}=\frac{d}{d\, ct}\tilde{R}V=0.\label{eq:app2}\end{equation}

\noindent \textbf{Proof:}

\noindent In the following, we do not consider explicitly the transverse
coordinates (to be considered constant). The variables of interest
are the time $ct\equiv x^{0}$ and the longitudinal coordinate $x^{\Vert}\equiv z$.
We use for any function $f(t,z)$ the notation $\partial^{0}f=\partial f/\partial\, ct$,
$\partial^{z}f=-\partial f/\partial z$. For the {}``total'' derivatives,
we use $d^{0}f=df/d\, ct$, $d^{z}f=-df/dz$.

We consider for given fixed $h$ the retarded time $t^{*}(t,z-h)$.
The retarded time corresponding to a critical time is given as\begin{equation}
t_{k}^{*}=t^{*}(t_{k},z-h),\end{equation}
with $t_{k}=t_{k}(z-h)$. So we have $t_{k}^{*}=t_{k}^{*}(z-h)$.
We have

\begin{equation}
d^{z}t_{k}^{*}=\partial^{0}ct^{*}\, d^{z}t_{k}+\partial^{z}t^{*}.\end{equation}
In the following we make extensively use of definitions and relations
from reference \citet{klaus1}. For $t=t_{k}$, we have \begin{equation}
\widetilde{R}V=c(t_{k}-t_{k}^{*})+\tilde{R}_{j}V^{j}|_{t^{*}=t_{k}^{*}}=0.\end{equation}
We compute the derivative $d^{z}$ :\begin{eqnarray}
 &  & \:0\:=d^{z}ct_{k}-d^{z}ct_{k}^{*}+(\bar{g}_{\: j}^{z}-\tilde{V}_{j}d^{z}ct_{k}^{*})V^{j}\\
 &  & \quad=d^{z}ct_{k}-(1-\tilde{V}^{j}V^{j})d^{z}ct_{k}^{*}+\bar{V}^{z}\\
 &  & \quad=d^{z}ct_{k}-\tilde{V}Vd^{z}ct_{k}^{*}+\bar{V}^{z}\\
 &  & \quad=d^{z}ct_{k}-\tilde{V}V(\partial^{0}ct^{*}\, d^{z}ct_{k}+\partial^{z}ct^{*})+\bar{V}^{z},\end{eqnarray}
which leads to (using $\bar{V}^{z}=\bar{V}^{\Vert}$) \begin{equation}
d^{z}ct_{k}=\frac{-\bar{V}^{\Vert}+\tilde{V}V\,\partial^{z}ct^{*}}{1-\tilde{V}V\,\partial^{0}ct^{*}}.\end{equation}
Using \begin{equation}
\partial^{\alpha}ct^{*}=\frac{\bar{R}^{\alpha}}{\widetilde{R}V},\end{equation}
we get \begin{equation}
d^{z}ct_{k}=-\frac{\bar{V}^{\Vert}-\tilde{V}V\,\bar{R}^{\Vert}/\widetilde{R}V}{1-\tilde{V}V\,\bar{R}^{0}/\widetilde{R}V}.\end{equation}
The time derivative of $h_{k}$ as obtained from its definition is\begin{equation}
d^{0}h_{k}=\left(d^{z}\, ct_{k}\right)^{-1},\end{equation}
which gives\begin{equation}
d^{0}h_{k}=-\frac{\widetilde{R}V-\tilde{V}V\,\bar{R}^{0}}{\widetilde{R}V\,\bar{V}^{\Vert}-\tilde{V}V\,\bar{R}^{\Vert}}.\end{equation}
Using $\widetilde{R}V=0$, and $\bar{R}^{\Vert}\approx L=\bar{R}^{0}$,
we find\begin{equation}
d^{0}h_{k}=-1.\end{equation}
 The other derivative of $h_{k}$ is trivial: \begin{equation}
d^{z}h_{k}=-1.\end{equation}

The potential $A$, its denominator $\tilde{R}V$, and also the argument
$ct^{*}$of its numerator $J$ are evaluated at the parallel coordinate
$x^{\Vert}-h_{k}+\lambda$, so the total time derivatives are \begin{equation}
d^{0}=\partial^{0}+d^{0}h_{k}\,\partial^{z},\end{equation}
explicitly\begin{equation}
d^{0}=\partial^{0}-\partial^{z}.\label{eq:ddct}\end{equation}
We get\begin{equation}
d^{0}ct^{*}=(\partial^{0}-\partial^{z})ct^{*}=\frac{\bar{R}^{0}-\bar{R}^{z}}{\widetilde{R}V}=0,\end{equation}
and with \begin{equation}
\partial^{\alpha}\tilde{R}V=\bar{V}^{\alpha}-\tilde{V}V\,\partial^{\alpha}ct^{*}\end{equation}
(eq. (22) from \citet{klaus1}), we obtain\begin{equation}
d^{0}\tilde{R}V=0.\end{equation}

\end{document}